\documentstyle[aps]{revtex}
\begin{document}
\draft
\preprint{\vbox{\hbox{DOE/ER/40762-200}\hbox{UMD PP\#00-045}}}
\title{Symmetries of Excited Heavy Baryons In The Heavy Quark And Large 
{\boldmath $N_c$} Limit}
\author{Chi--Keung Chow and Thomas D.~Cohen}
\address{Department of Physics, University of Maryland, College Park, 
20742-4111.}
\date{\today}
\maketitle
\begin{abstract} 
We demonstrate in a model independent way that, in the combined heavy quark 
and large $N_c$ limit, there exists a new contracted U(4) symmetry which 
connects orbitally excited heavy baryons to the ground states.  
\end{abstract}

Due to our inability to solve nonperturbative QCD from first principles, 
most of our quantitative understanding of low energy hadron properties 
are based on symmetry considerations.  
The most notable of these schemes is chiral perturbation theory, which is 
based on the fact that the QCD Lagrangian is approximately chirally invariant. 
On the other hand, there are emergent symmetries which are {\it not\/} 
symmetries (not even approximate symmetries) of the QCD Lagrangian, but 
emerge as symmetries of an effective theory obtained by taking certain 
limits.  
Two famous examples of such emergent symmetries, namely the heavy quark 
symmetry \cite{HQ} and the large $N_c$ spin-flavor symmetry \cite{LN}, have 
important phenomenological implications and are well discussed in the 
literature.  

In this paper, we will discuss a new emergent symmetry of QCD, which emerges 
in the heavy baryon (baryon with a single heavy quark) sector in the 
combined heavy quark and large $N_c$ limit.   
As we will see below, this contracted U(4) symmetry connects the ground state 
baryon to some of its orbitally excited states.  
As a result, static properties like the axial current couplings and the 
moments of the weak form factors of these orbitally excited states can be 
related to their counterparts of the ground state.  
While some of these results have been discussed before in the literature, 
this is the first time where they are presented as symmetry predictions.  
Moreover, unlike previous studies, the analysis here is essentially model 
independent, depending only on the heavy quark and large $N_c$ limits.  
We will outline the steps through which the existence of this symmetry can be 
demonstrated, while the details of the construction will be reported in the 
longer paper \cite{next}.  

It has been pointed out that, in the combined heavy quark and large $N_c$ 
limit, a heavy baryon can be regarded as a bound state of a heavy meson 
and a light baryon (a baryon for which all valence quarks are light) 
\cite{genius}.  
(Similar models are studied in \cite{others}.)  
For concreteness, we will always adopt the prescription that the 
heavy quark limit is taken before the large $N_c$ limit.  
Since both constituents are infinitely massive (the heavy meson in the 
heavy quark limit, the light baryon in the large $N_c$ limit), a small 
attraction between them is sufficient to ensure the existence of a bound 
state.  
By the usual large $N_c$ counting rule it can be shown that the binding 
potential is of order $N_c^0$.  
Moreover, as the kinetic energy term is suppressed by the large reduced 
mass of the bound state, the wave function does not spread and is instead 
localized at the bottom of the potential.  
As a result, it can be approximated by a simple harmonic potential: 
$V(x) = V_0 + {1\over2} \kappa \vec x^2$.  
By the large $N_c$ counting rules, both $V_0$ and $\kappa$ are of order 
$N_c^0$, and it has been shown that in the model studied in 
Refs.~\cite{genius}, $V_0<0$, $\kappa>0$, {\it i.e.}, the potential is 
attractive and can support bound states.    
When the bound state is the ground state of the simple harmonic oscillator, 
it is a $\Lambda_Q$, the lightest heavy baryon containing the heavy quark $Q$.
On the other hand, excited states in the simple harmonic oscillator are
orbitally excited heavy baryons.  
We emphasize that this description of a heavy baryon is a model, which 
is {\it not\/} directly related to QCD.  

After describing the physical picture of this model, which we will refer to 
as the bound state picture, we make the crucial observation that the 
excitation energy $\omega = \sqrt{\kappa/\mu}$ is small, where $\mu$ 
is the reduced mass of the bound state.  
By first taking the heavy quark limit, $\mu = m_N$ (mass of the light 
baryon) scales like $N_c$.  
Since the spring constant $\kappa$ is of order $N_c^0$, $\omega$ scales 
like $N_c^{-1/2}$ and vanishes in the large $N_c$ limit.  
This implies that when $N_c\to\infty$, the whole tower of excited states 
become degenerate with the ground state --- a classic signature of an 
emergent symmetry.  

What is the symmetry group of this emergent symmetry then?  
It has to contain, as a subgroup, the symmetry group of a three-dimensional 
simple harmonic oscillator, namely U(3) generated by $T_{ij} = a_i^\dag a_j$ 
($i,j = 1$, 2, 3) where $a_j$ is the annihilation operator in the $j$-th 
direction.  
These $T_{ij}$'s satisfy the U(3) commutation relations.  
\begin{equation}
[T_{ij},T_{kl}] = \delta_{il} T_{kj} + \delta_{kj} T_{il}.  
\label{com}
\end{equation}
When $N_c \to\infty$ and the excited states become degenerate with the 
ground state, the annihilation and creation operators $a_j$ and $a_i^\dag$ 
($i,j = 1$, 2, 3) also become generators of the emergent symmetry.  
The additional commutation relations are 
\begin{equation} 
[a_j, T_{kl}] = \delta_{kj} a_l, \quad [a_i^\dag, T_{kl}] = -\delta_{il} 
a_k^\dag, \quad [a_j, a_i^\dag] = \delta_{ij} {\bf 1}, 
\end{equation}
where {\bf 1} is the identity operator.  
These sixteen generators $\{T_{ij}, a_l, a_k^\dag, {\bf 1}\}$ form the 
{\it spectrum generating algebra\/} of a three-dimensional harmonic 
oscillator.  
It is related to the usual U(4) algebra, generated by $T_{ij}$ ($i,j=1$, 2, 
3, 4) satisfying commutation relations (\ref{com}) by the following 
limiting procedure: 
\begin{equation}
a_j = \lim_{R\to\infty} T_{4j}/R, \quad a_i^\dag = \lim_{R\to\infty} T_{i4}/R,
\quad {\bf 1} = \lim_{R\to\infty} T_{44}/R^2.
\end{equation}
Such a limiting procedure is called a group contraction, and hence 
the group generated by $\{T_{ij}, a_l, a_k^\dag, {\bf 1}\}$ is called a 
{\it contracted\/} U(4) group.  

So we have shown that the contracted U(4) is a symmetry of the bound state 
picture.  
But is it also a symmetry of QCD itself?  
We claim the answer is affirmative, and it is the aim of this paper to 
demonstrate the existence of this U(4) symmetry in QCD itself.  
We will first construct operators $p_j$ and $x_j$ ($j=1$, 2, 3), which 
correspond to the momentum and central position of the ``brown muck'' (light 
degrees of freedom) of the heavy baryon.  
From them operators $a_j$ and $a_j^\dag$ are constructed.  
By considering the double and triple commutator of the QCD Hamiltonian 
$\cal H$ with these operators, one can show that $\cal H$ is at most a 
bilinear in $a_j$ and $a_j^\dag$ as $N_c\to\infty$. 
Then it is straightforward to show that in the large $N_c$ limit, 
\begin{equation}
{\cal H} = H+H', \qquad H=\omega \sum_{j=1}^3 a_j^\dag a_j, 
\label{H}
\end{equation}
where $\omega$ is a parameter of order $N_c^{-1/2}$ and $H'$ is an operator 
which commutes with all the $a_j$ and $a_j^\dag$. 
This Hamiltonian clearly has a contracted U(4) emergent symmetry in the large 
$N_c$ limit as $\omega\to0$.  

Due to the conservation of baryon number and heavy quark number (in the heavy 
quark limit), it is well-defined to restrict our attention to the {\it heavy 
baryon Hilbert space}, {\it i.e.}, the subspace with both heavy quark number 
and baryon number equal to unity.  
In this subspace, we will define operators which correspond to the momenta 
and positions of the heavy quark and brown muck of the heavy baryon.  
So let $\cal H$ be the QCD Hamiltonian in this heavy baryon Hilbert space, 
$P_j$ ($j=1$, 2, 3) be the momentum operators, and $X_j$ be the operators 
conjugate to $P_j$.  
On the other hand, in the heavy quark limit, both the heavy quark mass $m_Q$ 
and the heavy quark momentum operators ${P_Q}_j$ ($j=1$, 2, 3) are 
well-defined up to order $N_c^0$ ambiguities, and ${X_Q}_j$ are the operators 
conjugate to ${P_Q}_j$.  
These operators satisfy the following operator identities:  
\begin{equation}
[{\cal H}X_j,{\cal H}]=iP_j,\qquad [m_Q{X_Q}_j, {\cal H}]=i{P_Q}_j, 
\label{oi}
\end{equation}
where the first identity follows from Poincare invariance, and the second from 
heavy quark symmetry where $\Lambda_{QCD}/m_Q$ corrections are dropped 
\cite{HQ}.  
Note that in the heavy quark limit, both $X_j$ and ${X_Q}_j$ commute with 
$\cal H$ and hence are constants of motion.  
Lastly, we define the brown muck momentum operators $p_j=P_j-{P_Q}_j$, and 
$x_j$ to be their conjugate operators.  

The Hamiltonian $H$ can be decomposed into three pieces: ${\cal H}=m_Q+m_N+
\tilde H$, where by the large $N_c$ scaling rules $m_N\sim N_c$ and 
$\tilde H\sim N_c^0$. 
Moreover, since ${\cal H}X_j = m_Q{X_Q}_j + (m_N+\tilde H)x_j$, one can 
subtract the operator identities (\ref{oi}) to obtain 
\begin{equation}
[m_Nx_j,{\cal H}] = [{\cal H}X_j-m_Q{X_Q}_j,{\cal H}] = i(P_j-{P_Q}_j)
=ip_j, 
\end{equation}
with the term proportional to $\tilde H$ dropped as it is order $N_c^{-1}$ 
suppressed relative to the leading order.  
So one has 
\begin{equation}
[x_k, [x_j,{\cal H}]] = - \delta_{jk} /m_N \sim {\cal O}(N_c^{-1})
(1+{\cal O}(N_c^{-1})).  
\label{dc1}
\end{equation}
On the other hand, the double commutator $[p_k, [p_j,{\cal H}]]=[p_k, [p_j,
\tilde H]]$ measures the second order energy change when the heavy quark is 
spatially moved with respect to the brown muck.  
Since $\tilde H$ is of order $N_c^0$, the double commutator is generically 
also of the same order.  
(For later use, we will mention in passing that, by the same logic, all 
multiple commutators like $[p_i,[p_j,\dots [p_k,{\cal H}]\dots]]$ are also of 
order $N_c^0$.)  
We will define 
\begin{equation}
\hat\kappa = - [p_k, [p_j,{\cal H}]] \sim{\cal O}(N_c^{0})
(1+{\cal O}(N_c^{-1})), \qquad \kappa = \langle G|\hat\kappa|G\rangle
\label{dc2}
\end{equation}
where $|G\rangle$ is the ground state of QCD Hamiltonian $\cal H$.  

Now let us take the heavy quark limit, and without loss of generality, set 
the constants of motion $X_j={X_Q}_j=0$, so that the heavy baryon is sitting 
at the origin, and $x_j$ becomes the position of the center of the brown muck 
relative to the heavy quark, which is the center of mass of the heavy baryon.  
Then note that both Eqs.~(\ref{dc1}) and (\ref{dc2}) will still hold if the 
QCD Hamiltonian $\cal H$ in the double commutators are replaced by $H$, the 
Hamiltonian of a simple harmonic oscillator.  
\begin{equation}
H= \sum_{j=1}^3 \left[{(p_j)^2\over 2m_N} + {\kappa(x_j)^2\over2} - 
{\omega\over 2}\right] = \omega \sum_{j=1}^3 a_j^\dag a_j, \qquad 
a_j = \sqrt{m_N\omega\over2}x_j + i\sqrt{1\over2m_N\omega}p_j, 
\end{equation}
where $a_j^\dag$ is the hermitian conjugate of $a_j$ and 
$\omega=\sqrt{\kappa/m_N}\sim {\cal O}(N_c^{-1/2})$.  
The contracted U(4) symmetry mentioned above is precisely the spectrum 
generating algebra of $H$ and becomes an emergent symmetry as $\omega\to 0$ 
in the large $N_c$ limit.  
On the other hand, to demonstrate that this contracted U(4) is a symmetry of 
QCD, one needs to show that the generators of the contracted U(4) commute with 
the QCD Hamiltonian $\cal H$, or equivalently, show that ${\cal H} = H+H'$, 
where $H'$ commutes with $a_j$ and $a_j^\dag$ in the large $N_c$ limit, 
{\it i.e.}, $[a_j,H']=[a_j^\dag,H']=0$.  

Our strategy is to study all possible double and triple commutators of 
$\cal H$ with $a_j$ and $a_j^\dag$.  
It is straightforward to show that the vanishing of the triple commutators 
$[a_i,[a_j,[a_k,{\cal H}]]]$ and $[a_i^\dag,[a_j,[a_k,{\cal H}]]]$ implies 
that $[a_k,{\cal H}] = Ca_k+Da_k^\dag$, where both $C$ and $D$ commute with 
both $a_k$ and $a_k^\dag$.  
(A possible constant term is ruled out by parity.)  
But then ${\cal H}$ is at most a bilinear in $a_k$ and $a_k^\dag$: 
\begin{equation}
{\cal H}=\sum_{k=1}^3 Ca_k^\dag a_k + D(a_k a_k + a_k^\dag a_k^\dag) +H', 
\label{form}
\end{equation}
where $H'$ commutes with all the $a_k$ and $a_k^\dag$.  
Moreover, the forms of the operators $C$ and $D$ can be fixed by the relations 
$C\delta_{jk}=[a_j^\dag,[a_k,{\cal H}]]$ and $D\delta_{jk}=[a_j,[a_k,{\cal H}
]]$. 
Hence if $C=\omega$ and $D=0$, ${\cal H} = H+H'$ with both terms invariant 
under the contracted U(4) symmetry, completing the argument that this is a 
symmetry of QCD.  

The triple commutators $[a_i,[a_j,[a_k,{\cal H}]]]$ and $[a_i^\dag,[a_j,[a_k,
{\cal H}]]]$ are both linear combinations (with coefficients ${\cal O}
(N_c^0)$) of these four terms: 
\begin{eqnarray}
T^{(0)} &= (m_N\omega)^{-3/2} [p_i,[p_j,[p_k,{\cal H}]]],\qquad
T^{(1)} &= (m_N\omega)^{-1/2} [p_i,[p_j,[x_k,{\cal H}]]],\nonumber\\
T^{(2)} &= (m_N\omega)^{1/2} [p_i,[x_j,[x_k,{\cal H}]]],\qquad\;\;
T^{(3)} &= (m_N\omega)^{3/2} [x_i,[x_j,[x_k,{\cal H}]]].
\end{eqnarray}
As mentioned before, the triple $p$ commutator is at most of order 
$N_c^0$, and hence $T^{(0)}\sim {\cal O}(N_c^{-3/4})$.  
All of the other three triple commutators are also small as  
$[x_k,{\cal H}] = ip_k/m_N + {\cal O}(N_c^{-2})$, and the first term does 
not contribute to the triple commutators.  
So $T^{(1)}\sim N_c^{-9/4}$, $T^{(2)}\sim N_c^{-7/4}$ and $T^{(3)}\sim 
N_c^{-5/4}$.  
All four terms are smaller than $\omega\sim N_c^{-1/2}$ in the large $N_c$ 
limit, and hence so are the triple commutators $[a_i,[a_j,[a_k,{\cal H}]]]$ 
and $[a_i^\dag,[a_j,[a_k,{\cal H}]]]$.  
By the strategy outlined above, the vanishings of these triple commutators 
in the large $N_c$ limit imply that $\cal H$ is of the form in 
Eq.~(\ref{form}).  

Lastly, we want to show that $C=\omega$ and $D=0$, which is equivalent to 
showing that $[x_j,[x_j,{\cal H}]]=-1/m_N$ and $[p_j,[p_j,{\cal H}]]=
-\kappa$.  
(The index $j$ is not summed over.)  
While the former is true from Eq.~(\ref{dc1}), the latter is not: from 
Eq.~(\ref{dc2}) the double $p$ commutator is $\hat\kappa$, which is in general 
not identical with its ground state expectation value $\kappa$.  
However, it is true for states in the {\it ground state band}, which is the 
subspace spanned by states of the form $(a_x^\dag)^{n_x} (a_y^\dag)^{n_y} 
(a_z^\dag)^{n_z}|G\rangle$.  
So we conclude that, in the ground state band, $\cal H$ does have the form 
stated in Eq.~(\ref{H}), and hence is invariant under the contracted U(4) 
group in the large $N_c$ limit. 
Note that this symmetry, like the familiar large $N_c$ spin-flavor symmetry 
\cite{LN}, only applies to a particular subspace of the theory --- in this 
case the ground state band.  

So we have demonstrated that this contracted U(4) symmetry is not only a 
symmetry of the bound state picture, but indeed a symmetry of QCD.  
Near the combined heavy quark and large $N_c$ limit, there exists a band of 
low-lying heavy baryons, labeled by $n_x$, $n_y$ and $n_z$, the number of 
excitation quanta in the $x$, $y$ and $z$ directions, and the excitation 
energies are $(n_x+n_y+n_z)\omega$. 
As $N_c\to\infty$, $\omega\to0$ and the whole band become degenerate.  

Such a symmetry has interesting phenomenological implications.  
For example, consider light quark form factors of the form $J_\ell=
\bar q \Gamma q$, where $\Gamma$ is an arbitrary combination of the gamma 
matrices, with momentum transfer of order $N_c\Lambda_{QCD}$.    
(In general the $\bar q$ and $q$ may have different flavors.)  
Since $J_\ell$ is a single quark operator while the U(4) generators 
are collective coordinates involving $N_c$ quarks, it follows that the 
commutators like $[a_j, J_\ell]$ are of order ${\cal O}(N_c^{-1})$ 
and vanish as $N_c\to\infty$.  
As a result, the light quark form factor is diagonal in the large $N_c$ limit. 
\begin{equation}
\langle \Lambda_Q^{(n_x',n_y',n_z')}(p')|J_\ell|\Lambda_Q^{(n_x,n_y,n_z)}(p)
\rangle = \delta_{n_xn_x'} \delta_{n_yn_y'} \delta_{n_zn_z'} f(q^2), \qquad
q^2 = (p-p')^2 \sim N_c^0 \Lambda_{QCD}, 
\end{equation}
where $F(q^2)$ is an order $N_c$ form factor, which is a function of $q^2$, 
the square of the momentum transfer.  
Consequently, $g_{\pi\Lambda_Q\Lambda_Q}\sim F(q^2)/f_\pi\sim N_c^{1/2}$, 
as the pion decay constant $f_\pi\sim N_c^{1/2}$.  
On the other hand, $g_{\pi\Lambda_Q\Lambda_Q^*}$ is suppressed in the large 
$N_c$ limit, a result which has been discussed in \cite{CW} in the context 
of the bound state picture, but here presented as an implication of the 
contracted U(4) group theory.  

Like the light quark currents, heavy quark currents $J_h=\bar Q' \Gamma Q$ is 
also invariant under the contracted U(4) symmetry.  
Naively, one may conclude that the Isgur-Wise form factors, which are the 
matrix elements of such heavy quark currents, also only connect initial and 
final states with the same $(n_x, n_y, n_z)$.  
Such conclusions are erroneous, however, as the final state is boosted 
relative to the initial state, and hence 
\begin{equation}
\eta(w) \equiv \langle \Lambda_{Q'}^{(n_x',n_y',n_z')}(v')|J_h|
\Lambda_Q^{(n_x,n_y,n_z)}(v)\rangle = \langle \Lambda_{Q'}^{(n_x',n_y',n_z')}
(v)|B^\dag_{v-v'} J_h|\Lambda_Q^{(n_x,n_y,n_z)}(v)\rangle, 
\end{equation}
where $w$ is the scalar product of the four-velocities $v$ and $v'$, which 
is related to three-velocities $v_j$ and $v_j'$ by $w=1+|v_j-v_j'|^2/2$ in 
the non-relativistic limit.  
The boost operator $B_{v-v'}=\exp\left(i{\cal H}X_j(v-v')_j\right)$ 
boosts the heavy baryon from velocity $v$ to $v'$.  
In the large $N_c$ limit, ${\cal H}X_j = m_{Q'}{X_Q}_j+m_Nx_j$, where 
the first term commutes with $a_j$ and $a_j^\dag$ but the second term does 
not. 
As a result, $B_{v-v'}$ does not commute with $a_j$ and $a_j^\dag$, and the 
Isgur-Wise form factors between states with different $(n_x, n_y, n_z)$ do 
not vanish: 
\begin{equation}
\eta(w) = \langle n_x',n_y',n_z|\exp\left(im_N x_j (v-v')_j\right) |
n_x,n_y,n_z\rangle, 
\end{equation}
where $|n_x,n_y,n_z\rangle$ is the simple harmonic eigenstates, and 
$x_j = (a_j+a_j^\dag)/\sqrt{2m_N\omega}$. 
This is simply a group theoretical expression which only depends on two 
parameters: $m_N$ and $\omega$, which can be fixed by measuring the excitation 
energy of the first excited state to be around 330 MeV.  
All Isgur-Wise form factors (or more exactly, all their derivatives at the 
point of zero recoil $w=1$) between different initial and final heavy baryon 
states can be expressed as calculable functions of $m_N$ and $\omega$.  
For example, at the point of zero recoil ($v=v'$ and $w=1$), the boost 
operator reduces to an identity operator and the Isgur-Wise form factors are  
non-zero if and only if $(n_x,n_y,n_z)=(n_x',n_y',n_z')$; {\it i.e.}, the 
ground state $\Lambda_Q$ can only decay into a ground state $\Lambda_{Q'}$, 
a well-known prediction of heavy quark symmetry \cite{HQ}.  
When the velocity transfer is non-zero but small, $w=1+\epsilon$, the ground 
state $\Lambda_Q$ can decay into excited $\Lambda_{Q'}$.  
Since $x_j$ is linear in $a_j^\dag$, however, at order $\epsilon$ the 
only non-vanishing excited state Isgur-Wise form factor is that to the 
first excited state, and it saturates both the Bjorken and Voloshin sum rules 
\cite{CW}. 
This analysis of the Isgur-Wise form factors recasts the studies of 
Refs.~\cite{genius,CW} in a model independent way.  

In the above, we have ignored the spins and flavors of the heavy baryons.  
The inclusion of these quantum numbers does not change the above analysis.  
In particular, both the spin-flavor symmetry for the brown muck and the 
flavor symmetry for the heavy quark, being generated by one-quark operators, 
commute with our contracted U(4) in the large $N_c$ limit.  
These extra excitation modes live in the $H'$ term in Eq.~(\ref{H}).  
For example, with two light flavors, $H'=\sigma I^2$, with $\sigma\sim 
N_c^{-1}$ for states with $I+s_\ell=0$, where $I$ is isospin and $s_\ell$ is 
the spin of brown muck (excluding any possible orbital angular momentum) 
\cite{PY}.  
So the eigenstates of $H'$ is a tower of states labeled by $I=0,1,\dots$, 
where the states with $I=0$ and 1 are the $\Lambda_Q$ and $\Sigma_Q^{(*)}$, 
respectively.  
Since $\sigma\ll\omega$ in the large $N_c$ limit, inclusion of such a $H'$ 
splits each simple harmonic eigenstate of $H$ into a whole tower of $I=s_\ell$ 
states.    

In conclusion, we have demonstrated that in the combined heavy quark and large 
$N_c$ limit, there exists a new emergent symmetry which connects orbitally 
excited states to the ground states.  
While such a symmetry has interesting phenomenological implications, its  
utility for quantitative predictions may be limited by potentially large 
corrections which are typically of order $N_c^{-1/2}$.  
For example, the effect of an anharmonic term $ax^4$ ($a\sim N_c^0$) leads to 
order $N_c^{-1/2}$ mixings between the simple harmonic oscillator states.  
However, even though the corrections may be large, the existence of such a 
symmetry provides a useful organization principle for low energy properties 
of heavy baryons, and may provide qualitative or semi-quantitative 
predictions.  

\bigskip

This work is supported by the U.S.~Department of Energy grant 
DE-FG02-93ER-40762.

\end{document}